\documentclass[ reprint, aps, prl, two column, showkeys, groupedaddress]{revtex4-1}
\usepackage{graphicx,epsfig}
\usepackage{amssymb}
\usepackage{amsmath}
\usepackage{bm}
\usepackage{textcomp}
\usepackage{soul,xcolor}

\usepackage{color}

\begin{document}
\setcounter{page}{1}
\setstcolor{red}

\title[]{Working principles of doping-well structures for high-mobility two-dimensional electron systems}
\author{Yoon Jang \surname{Chung}}
\author{K. A. \surname{Villegas Rosales}}
\author{K. W. \surname{Baldwin}}
\author{K. W. \surname{West}}
\author{M. \surname{Shayegan}}
\author{L. N. \surname{Pfeiffer}}
\affiliation{Department of Electrical Engineering, Princeton University, Princeton, NJ 08544, USA  }
\date{\today}

\begin{abstract}

Suppressing electron scattering is essential to achieve high-mobility two-dimensional electron systems (2DESs) that are clean enough to probe exotic interaction-driven phenomena. In heterostructures it is common practice to utilize modulation doping, where the ionized dopants are physically separated from the 2DES channel. The doping-well structure augments modulation doping by providing additional screening for all types of charged impurities in the vicinity of the 2DES, which is necessary to achieve record-breaking samples. Despite its prevalence in the design of ultra-high-mobility 2DESs, the working principles of the doping-well structure have not been reported. Here we elaborate on the mechanics of electron transfer from doping wells to the 2DES, focusing on GaAs/AlGaAs samples grown by molecular beam epitaxy. Based on this understanding we demonstrate how structural parameters in the doping well can be varied to tune the properties of the 2DES.

\end{abstract}
\maketitle

Two-dimensional electron systems (2DESs) offer a unique platform to investigate many-body electron physics. This is because at low enough temperatures where the kinetic energy of the electrons in the 2DES is determined by the Fermi energy, the 2DES density can be tuned in an experimentally accessible range so that the Coulomb and/or exchange interactions become prevalent. Additionally, the kinetic energy of the 2DES can be quenched to even lower values by applying a perpendicular magnetic field which quantizes the 2DES energy into Landau levels, further strengthening the influence of electron-electron interactions. Indeed, many intriguing phenomena that cannot be explained in the single-particle picture have been observed in 2DESs, some notable examples being the fractional quantum Hall effect \cite{Tsui,A}, Wigner solid formation \cite{Wigner1,Wigner2,B}, stripe/nematic phases \cite{C,D,E,F}, negative compressibility \cite{NC1,NC2}, ferromagnetism \cite{ferro}, and correlated insulating behavior/superconductivity \cite{graph1,graph2}.

Not all 2DESs display such intriguing characteristics. A critical factor in determining the viability to exhibit many-body effects in 2DESs is the `cleanliness' or quality of the system. In epitaxially-grown 2DESs, a historic achievement in this regard was the invention of modulation doping, which separates the ionized dopant atoms from the 2DES \cite{modulationdoping}. This significantly suppresses the amount of scattering the electrons in the 2DES experience and thus enhances the 2DES quality, often quantified by the low-temperature mobility and the strength of many-body states. Of course, the number of residual background impurities must also be minimized to attain ultra-high sample quality, and continuous efforts are made to purify of the source material used in ultra-high vacuum deposition chambers \cite{Ga1,Ga2,Al1,Al2}.

For samples grown by molecular beam epitaxy (MBE), in addition to implementing modulation doping and having clean source materials, another structural modification, commonly coined as the `doping well', is generally used to achieve record-mobility 2DESs. Invented by our group in 1991 \cite{labnotes}, the doping-well structure (DWS) had an immediate impact on sample quality for both single-sided \cite{Ruidu} and double-sided-doped GaAs 2DESs \cite{Eisenstein}. The emergence of delicate and exotic many-body phases observed in these samples was a direct consequence of the DWS design scheme. Such a DWS improves standard modulation doping by providing extra screening for both the residual and intentional dopant impurities in the sample, significantly reducing electron scattering events in the 2DES \cite{DW1,DW2}. Despite its widespread use since its inception to achieve ultra-high-quality 2DESs, the working principles of the DWS are still rather obscure. In this work, we provide a detailed explanation of the electron transfer process in GaAs/AlGaAs samples employing the DWS, and demonstrate how it can be controlled by changing the structural parameters.

\begin{figure} [t]
\centering
    \includegraphics[width=.47\textwidth]{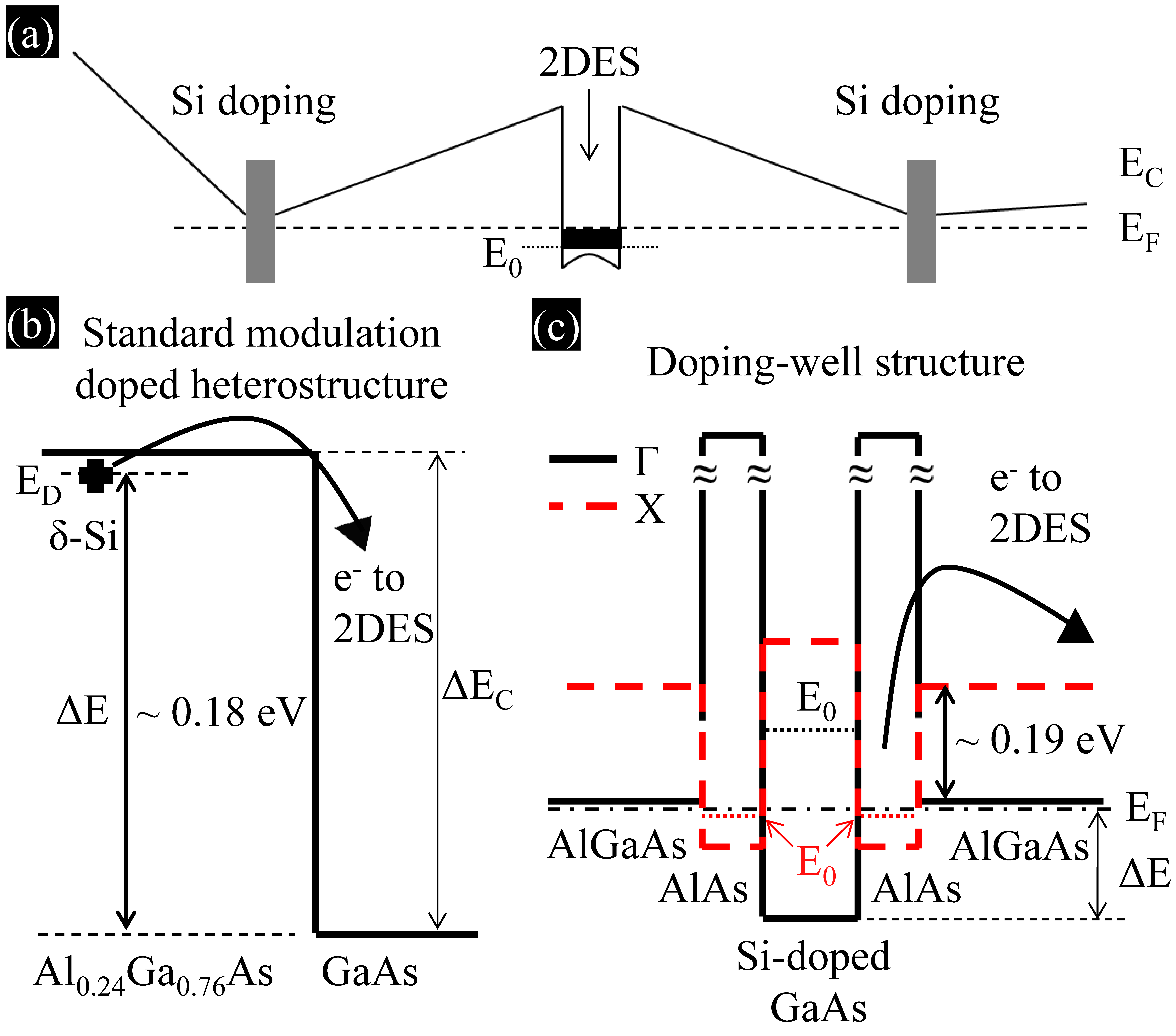}

 \caption{\label{fig1} (a) Schematic conduction-band diagram for a 2DES prepared in a GaAs quantum well flanked by AlGaAs barriers. Si acts as an electron donor and provides electrons to the 2DES channel via modulation doping. (b) Schematic conduction-band diagram for a standard modulation-doped heterostructure. Donor electrons transfer from the AlGaAs barrier to the GaAs 2DES channel because of the conduction-band offset between the two materials. (c) Schematic conduction-band diagram for a DWS with a barrier Al alloy fraction of 24\%. The $\Gamma$-band is denoted by the solid black line and the X-band by the dashed red line. The ground state energy $E_0$ of the doping well is shown in black and that of the cladding well is shown in red. Here, the donated electrons transfer to lower points of energy in the structure including the AlAs cladding quantum wells and the GaAs 2DES channel (not shown) because the strong confinement of the narrow wells pushes $E_0$ of the doping well up to high values.}
\end{figure}

It is useful to review the standard modulation doping process before describing how the DWS provides electrons to the 2DES. Figure 1(a) shows a schematic conduction-band diagram in the vicinity of a typical GaAs/AlGaAs 2DES. In this 2DES, Si is the most common dopant of choice, and as illustrated in the figure, the dopants are placed in the gray regions which are significantly distanced from the 2DES itself. The design of the doped region is where the standard modulation doping and DWSs differ. 
As shown in Fig. 1(b) \cite{flatband}, for standard modulation doping the energy difference $\Delta E$ between the dopant energy in the AlGaAs barrier and the ground state of the GaAs quantum well is the primary driving force for electron transfer. The density of the 2DES is related to $\Delta E$ by $n\simeq2({\Delta E}{\epsilon_b}/{se^2})$ assuming double-sided doping; here $s$ is the spacer thickness, $e$ is the fundamental electron charge, and $\epsilon_b$ is the dielectric constant of the barrier \cite{fnote1,Davies,Design}. Since Si acts as a hydrogenic donor in AlGaAs \cite{fnote1}, the 2DES density is then mostly determined by the conduction-band offset between AlGaAs and GaAs in the standard modulation-doped structure. 

In a DWS, the dopants are placed in a quantum well with a narrow well width rather than in the barrier. Instead of utilizing the conduction-band offset between two different materials, the DWS takes advantage of the large difference between the confinement energies in the narrow and wide quantum wells to transfer electrons to the 2DES channel. Although this energy difference alone is enough to push donor electrons toward the 2DES channel, in ultra-high-quality samples the doping wells have additional layers grown on their flanks as shown in Fig. 1(c). 

As depicted in Fig. 1(c) \cite{flatband}, for the case of GaAs/AlGaAs 2DESs, narrow AlAs wells are grown directly adjacent to the GaAs doping well. These `cladding wells' are beneficial because they host electrons that can screen the electric fields emanating from the ionized dopants in the doping well and other residual impurities \cite{DW1,DW2}. It is necessary for the doping and cladding wells to have their respective conduction-band minima at different places in the Brillouin zone, as this enables the ground state energy of the doping well to be higher than that of the cladding wells. Figure 1(c) shows that this condition is met for the GaAs/AlGaAs system since the conduction-band minimum is the $\Gamma$-point for GaAs while it is the X-point for AlAs. 

\begin{figure}[t]
 
 \centering
    \includegraphics[width=.47\textwidth]{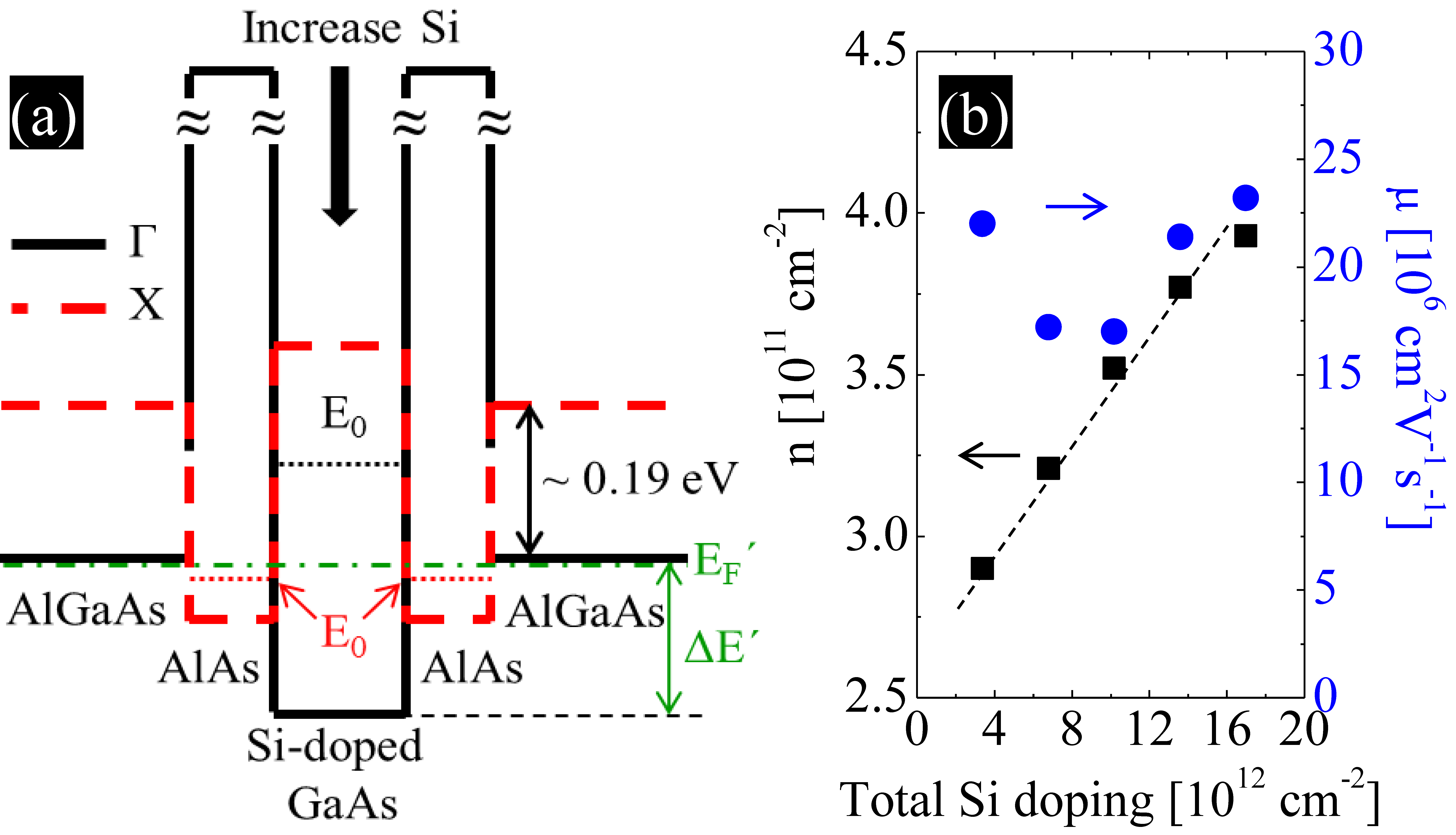} 
  \caption{\label{fig2} (a) Schematic conduction-band diagram of the DWS with a barrier Al alloy fraction of 24\% when the dopant concentration is increased. Here the Si doping density is the total doping density from both sides of the structure. Although the ground state energy ($E_0$) values stay the same, the increased concentration of Si increases the $E_F$ of the doping well to a slightly higher value of ${E_F}^\prime$. This in turn increases the density in the 2DES channel. (b) Measured 2DES density (black squares) and mobility (blue circles) in our 30 nm wide GaAs quantum well as a function of Si doping concentration in the doping well at $T\simeq0.3$ K. The dashed line denotes a linear fit to the black data points with a slope of $\sim0.008$ and an intercept of $\sim2.68\times10^{11}$ cm$^{-2}$ .}
\end{figure}

Following the structural description of the DWS, we now hypothesize on how electron transfer to the 2DES occurs. As mentioned earlier, the energy term that determines the 2DES density in a standard modulation-doped structure is the donor level in the barrier. In the simplest model, this is because the donor level is where the electrons that transfer to the main quantum well originate in the doped region. Comparing Figs. 1(b) and (c), it is then reasonable to assume that the analogous energy level for a DWS would be the quasi-Fermi energy ($E_F$) in the cladding well \cite{newerfootnote}, as that is the energy of the electrons leaving the doped region to populate the 2DES. In the following paragraphs we will validate this hypothesis by demonstrating changes in the 2DES density as we vary the properties of the doping and cladding wells.

The most direct way to tune $E_F$ in the cladding well is to simply vary the dopant concentration in the doping well. Illustrated in Fig. 2(a) is a specific case where the doping concentration is increased in the doping well compared to the case of Fig. 1(c). Increasing the dopant concentration in the doping well increases the concentration of electrons populating the cladding well, and we would expect $E_F$ to increase to a higher value ${E_F}^\prime$. Based on the model presented above, this should lead to an increase in the 2DES density. 

\begin{figure}[t]

\centering
    \includegraphics[width=.47\textwidth]{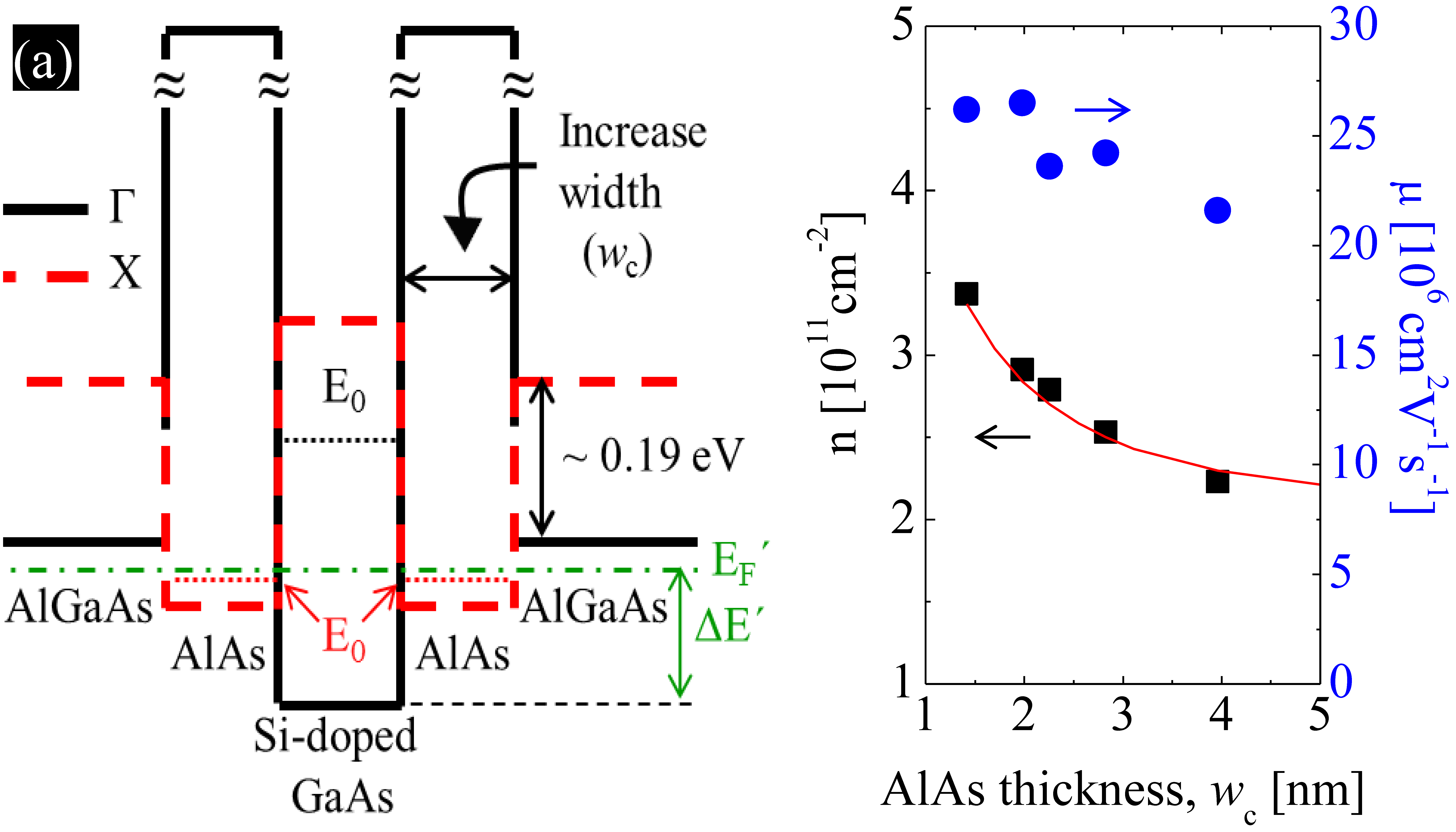} 
  \caption{\label{fig3} (a) Schematic conduction-band diagram of the DWS with a barrier Al alloy fraction of 24\% when the AlAs cladding well width is increased. The ground state energy ($E_0$) of the cladding wells decreases, decreasing $E_F$ of the doping well to a smaller value of ${E_F}^\prime$, for a fixed dopant concentration. This results in a decrease in the density of the 2DES channel. (b) Measured 2DES density (black squares) and mobility (blue circles) as a function of the AlAs cladding well width ($w_c$). The red line shows the expected 2DES density values derived from the calculated $E_0$ of the cladding wells.}
\end{figure} 

Figure 2(b) shows the measured 2DES density in GaAs/AlGaAs doping-well samples as the dopant concentration is varied. All samples are Si $\delta$-doped from both sides of the main 2DES GaAs channel and, except for their dopant concentration, have identical structures: an Al$_{0.24}$Ga$_{0.76}$As spacer of thickness 80 nm, doping well width of 2.8 nm, cladding well width of 2.0 nm, and main quantum well width of 30 nm. As shown in the figure, the 2DES density shows a roughly linear dependence on the Si concentration. A least square fit to the data yields a line with a slope of $\sim0.008$ and an intercept of $\sim2.68\times10^{11}$ cm$^{-2}$ . The slope can be interpreted as the Si doping efficiency, as it is a metric of how many electrons are generated in the main 2DES for each Si dopant atom. It is interesting to note that even including the electrons necessary to fill up the cladding wells when populating the main 2DES, the yield is still low, only $\simeq~0.22$ electrons/Si atom \cite{footnote}. It is possible that the majority of the electrons are trapped in the doping well itself, where a potential well may have formed due to the extremely high concentration  of positive charges from the ionized dopants \cite{delta1,delta2}. The intercept corresponds to the 2DES density expected of a structure where $\Delta E$ is solely determined by the confinement energy in the cladding well plus the conduction band offset between the GaAs $\Gamma$-band and AlAs X-band edges with no electrons populating the cladding wells.


The mobility values for the structures with different Si doping are also plotted in Fig. 2(b). The trend first exhibits a decline as the dopant concentration is increased and then shows an increase as more dopants are put into the system. The initial drop in mobility can be explained by two factors: the increased amount of scatterers present in the system, and the  contributions from the second subband. At the brink of occupying the second subband, the availability of the states to which electrons can scatter to suddenly increases and a decrease in mobility is expected. After this initial offset, the normal behavior of mobility increasing as the 2DES density increases is recovered once the second subband is substantially occupied \cite{secondsub}. We estimate from a self-consistent Schr\"{o}dinger-Poisson solver for a standard modulation-doped structure with an identical barrier alloy fraction ($x=0.24$) and main quantum well width (30 nm) that the electrons would start to occupy the second subband at a 2DES density of $\simeq3.1\times10^{11}$ cm$^{-2}$ in the structures we use. Considering the model mentioned above, it appears that the change in mobility we observe in Fig. 2(b) as the 2DES density increases is consistent with the second subband starting to be occupied around this density.



Another approach to adjust $E_F$ in the doped region is to modify the width of the AlAs cladding well ($w_c$), which would change its ground state energy ($E_0$ shown in red in Fig. 3(a)). For a fixed dopant concentration, this change should alter $E_F$. Such a situation is depicted in Fig. 3(a), where $w_c$ is increased in comparison to the GaAs/AlGaAs DWS illustrated in Fig. 1(c). Here the larger $w_c$ decreases the ground state energy of this cladding well and, assuming the same number of electrons populate the cladding well, $E_F$ should come down to a lower value ${E_F}^\prime$ when compared to a narrower cladding well. 

The measured 2DES density vs. $w_c$ data from our samples are presented in Fig. 3(b). The only structural variable for this series of samples was $w_c$, while all other factors such as spacer thickness, main quantum well width, and doping concentration were kept the same. The observed trend coincides quite well with our model. In fact, if we assume that the samples are only barely sufficiently doped and ${E_F}^\prime$ is roughly equal to $E_0$ in the cladding well, the data are in excellent agreement with the expected 2DES density calculated using simple, finite-potential-well calculations (red solid line in Fig. 3(b)) \cite{newfootnote}. The mobility values do not seem to significantly depend on $w_c$ other than the anticipated variance following the change in 2DES density ($\mu\sim n^{0.6}$) \cite{powerlaw}. 

Alternatively, Al$_x$Ga$_{1-x}$As cladding wells with varying $x$ can be used, instead of AlAs, as a means to change $E_F$ in the DWS. For example, increasing the Ga content in the cladding well raises its (X-point) conduction-band minimum. Then, even with the same $w_c$, dopant concentration, and spacer thickness conditions, $E_F$ would also be increased. Figure 4(a) portrays a schematic band diagram for this specific case, where a higher quasi-Fermi energy ${E_F}^\prime$ is drawn compared to the case of Fig. 1(c).

\begin{figure}[t]
\centering
    \includegraphics[width=.47\textwidth]{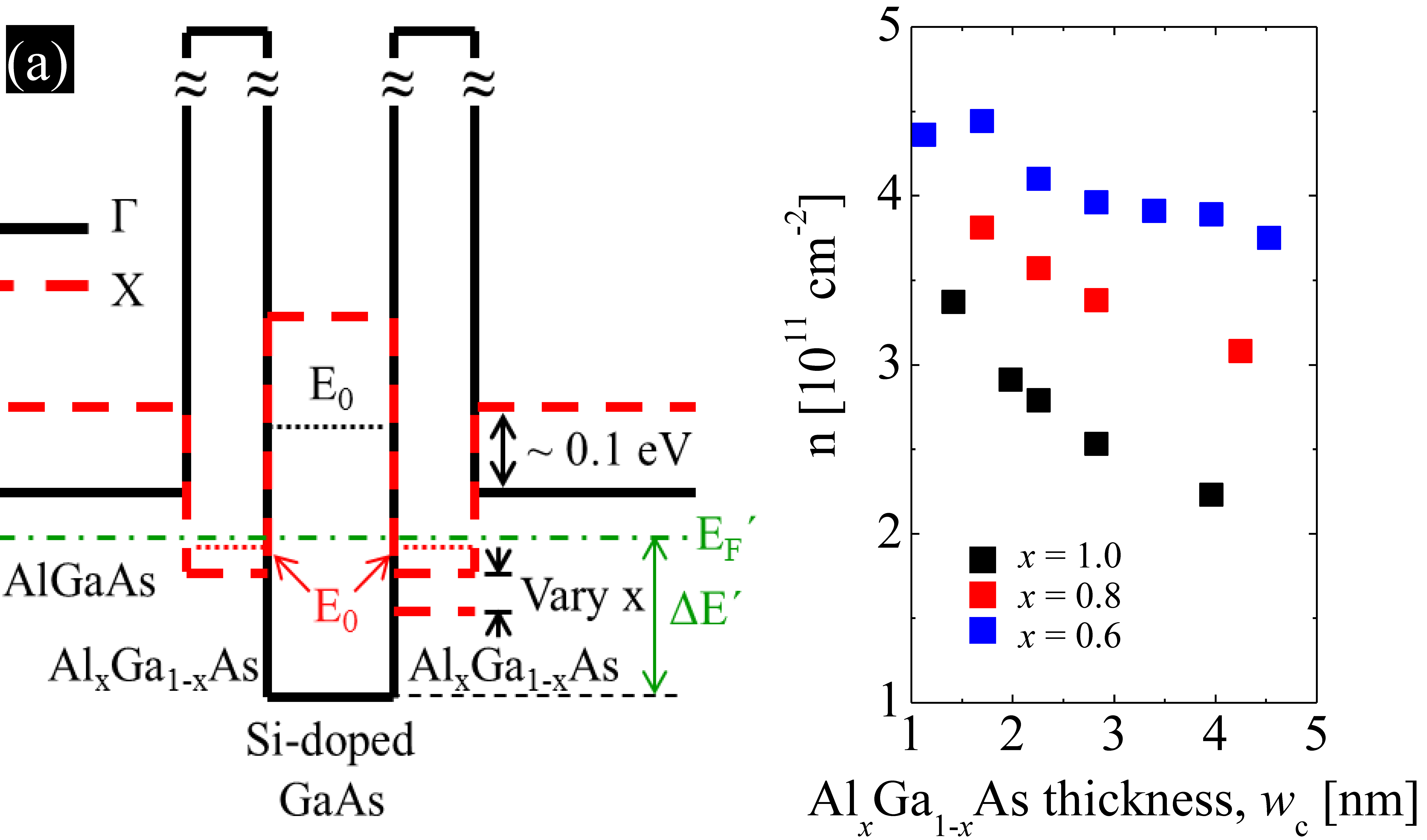} 

  \caption{\label{fig4} (a) Schematic conduction-band diagram of the doping-well structure with a barrier Al alloy fraction of 32\% when the alloy composition ($x$) of the cladding well is varied. Introducing an alloy modifies the conduction-band of the cladding well, changing $E_F$ of the doping well even when the dopant concentration and cladding well width are fixed. (b) Measured 2DES density as a function of Al$_x$Ga$_{1-x}$As cladding well width for Al compositions $x=1.0$ (black), $x= 0.8$ (red), and $x= 0.6$ (blue). }
\end{figure}

\begin{figure}[t]
\centering
    \includegraphics[width=.47\textwidth]{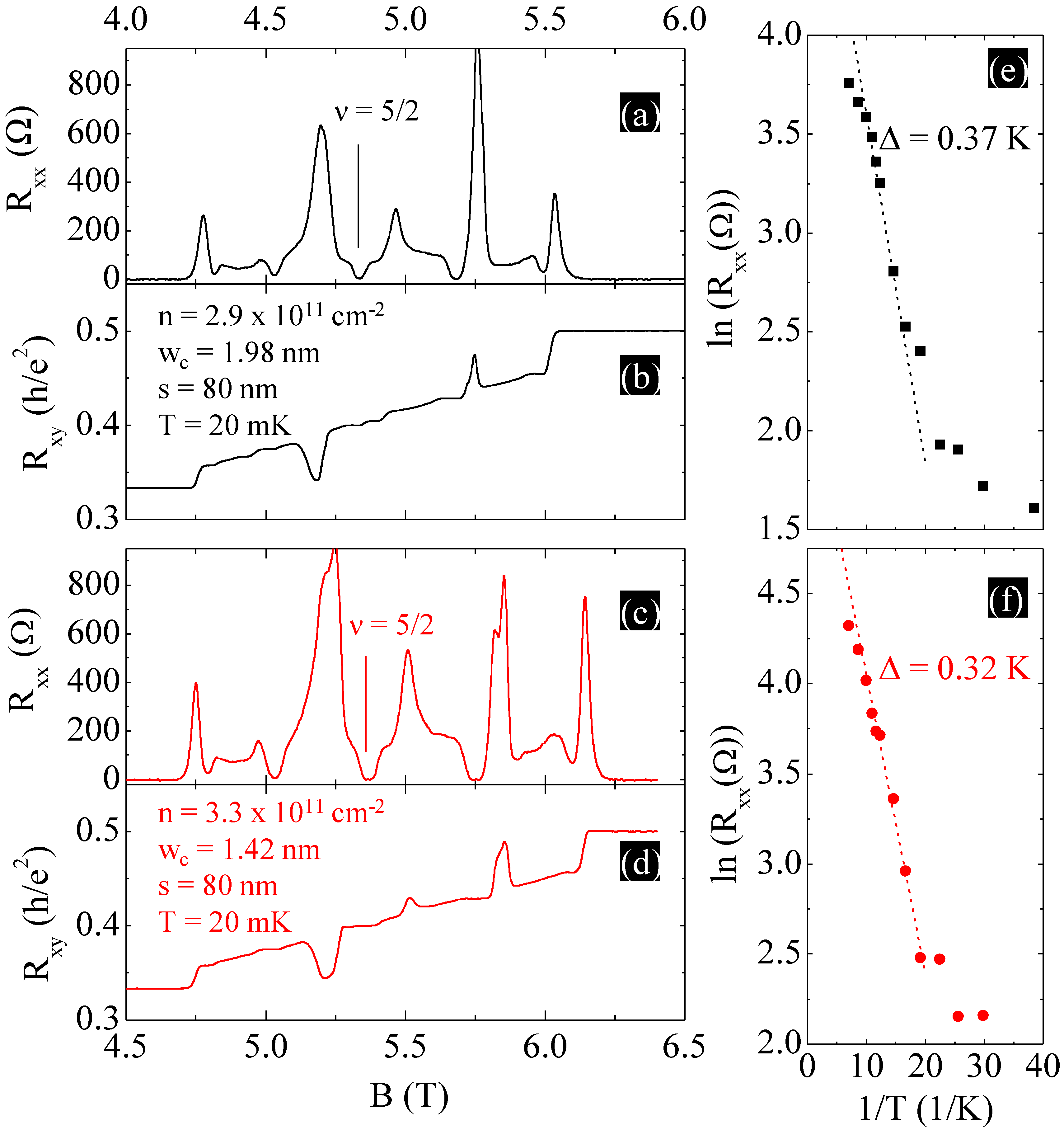} 

  \caption{\label{fig5} Representative $R_{xx}$ ((a), (c)), and Hall ((b), (d)) data in the vicinity of $\nu=5/2$ for GaAs 2DESs prepared using doping-well structures with cladding well widths of $w_c=1.98$ (black traces) and 1.42 nm (red traces). The spacer thickness is 80 nm for both samples. (e) and (f) show the temperature dependent $R_{xx}$ of the $\nu=5/2$ fractional quantum Hall state for the two samples. The dashed lines denote the linear fits used to extract the activation gap, $\Delta$. }
\end{figure}

We plot the measured 2DES density as a function of $w_c$ for various alloy compositions in Fig. 4(b). The black, red, and blue squares represent data for samples with Al composition ($x$) of 1.0, 0.8, and 0.6, respectively, in the Al$_x$Ga$_{1-x}$As cladding well. We would like to note that in some of these structures with smaller cladding alloy compositions and well widths, the barrier Al alloy fraction was adjusted to larger values to provide proper confinement for the cladding wells and prevent the formation of a parallel channel. For all $w_c$ values, it is clear that decreasing the Al composition results in an increased 2DES density. This is consistent with the fact that the X-band minimum in Al$_x$Ga$_{1-x}$As increases as $x$ is decreased \cite{Design,Adachi}. The decreasing 2DES density trend observed as $w_c$ increases also persists for all alloy compositions, confirming that the working principles for DWSs discussed earlier are generally applicable. Although it was difficult to observe a clear trend in mobility possibly due to factors such as the variation in barrier alloy fraction for some of the structures, all samples had mobility values in the vicinity of $\sim20\times10^6$ cm$^2$ V$^{-1}$s$^{-1}$ at $T\simeq0.3$ K.

Finally, it is worthwhile to comment on the quality of the 2DESs prepared by varying the parameters of the DWS. Figures 5 (a)-(d) show low-temperature ($T\simeq 20$ mK) longitudinal resistance ($R_{xx}$) and Hall ($R_{xy}$) data taken in the vicinity of the fractional quantum Hall state at Landau level filling $\nu=5/2$ for two different 2DESs with different doping well designs. The black traces in Figs. 5(a) and (b) are taken from a wafer with $w_c=1.98$ nm, while the red traces in Figs. 5(c) and (d) are from a sample with $w_c=1.42$ nm. In both cases the spacer thickness was 80 nm with a doping well thickness of 2.8 nm and the cladding wells were pure AlAs ($x=1.0$). The fractional quantum Hall and reentrant integer quantum Hall features observed in the data closely resemble those reported for extremely high quality 2DESs at similar temperatures \cite{Gabor}. We also measured the gap for the $\nu=5/2$ fractional quantum Hall state in these samples by analyzing the activated behavior of $R_{xx}$. Figures 5(e) and (f) show Arrhenius plots for the 2DESs with $w_c=1.98$ and 1.42 nm, where the measured energy gaps are $\Delta=0.37$ K and 0.32 K, respectively. These energy gap values are comparable to the best results in literature, and we believe they can be further improved by optimizing cooldown procedures such as light illumination \cite{Manfra}. All in all, our data clearly demonstrate that DWS parameters can be tuned to achieve very high quality samples over a range of densities even while keeping the spacer thickness fixed. This understanding provides multiple variables to further optimize quality for a given 2DES density in ultra-high-mobility samples. It may also be useful for the design and growth of samples that require a specific spacer thickness for its application, such as surface modulated devices where the 2DES must be placed a fixed distance away from the pattern.

In conclusion, we have established the working principles of the DWSs necessary for ultra high quality in epitaxially-grown 2DESs. The electron density in the 2DES channel can be tuned by varying the quasi Fermi energy in the DWS, which is analogous to the dopant energy level in standard modulation-doped structures. We demonstrate that this can be achieved by changing structural parameters such as dopant concentration, cladding well width, and cladding well alloy composition. Our data show that for DWS samples, the 2DES density can be modified over a wide range of values for a given spacer thickness while maintaining ultra high quality. Although in this work the GaAs/AlGaAs 2DES system is used as an example, the underlying physics of the DWS is quite general and can be applied to any epitaxially-grown 2DES with qualitatively similar conduction-band characteristics.

\begin{acknowledgments}
We acknowledge support through the NSF (Grants DMR 1709076 and ECCS 1906253) for measurements, and the NSF (Grant MRSEC DMR 1420541), the Gordon and Betty Moore Foundation (Grant GBMF4420), and the Department of Energy Basic Energy Sciences (Grant DEFG02-00-ER45841) for sample fabrication and characterization.
 \end{acknowledgments}


\begin{thebibliography}{99}

\bibitem{Tsui} D. C. Tsui, H. L. St\"{o}rmer, and A. C. Gossard, Two-dimensional magnetotransport in the extreme quantum limit. Phys. Rev. Lett. {\bf 48}, 1559 (1982).

\bibitem{A} J. K. Jain, \textit{Composite Fermions}. (Cambridge University Press, New York, 2007).

\bibitem{Wigner1} H. W. Jiang, R. L. Willett, H. L. St\"{o}rmer, D. C. Tsui, L. N. Pfeiffer, and K. W. West, Quantum liquid versus electron solid around $\nu=1/5$ Landau-level filling. Phys. Rev. Lett. {\bf 65}, 633 (1990).

\bibitem{Wigner2} V. J. Goldman, M. Santos, M. Shayegan, and J. E. Cunningham, Evidence for two-dimentional quantum Wigner crystal. Phys. Rev. Lett. {\bf 65}, 2189 (1990).

\bibitem{B} H. Deng, Y. Liu, I. Jo, L. N. Pfeiffer, K. W. West, K. W. Baldwin, and M. Shayegan, Commensurability oscillations of composite fermions induced by the periodic potential of a Wigner crystal. Phys. Rev. Lett. {\bf 117}, 096601 (2016).

\bibitem{C} M. P. Lilly, K. B. Cooper, J. P. Eisenstein, L. N. Pfeiffer, and K. W. West, Evidence for an anisotropic state of two-dimensional electrons in high Landau levels. Phys. Rev. Lett. {\bf 82}, 394 (1999).

\bibitem{D} R. R. Du, D. C. Tsui, H. L. St\"{o}rmer, L. N. Pfeiffer, and K. W. West, Strongly anisotropic transport in higher two-dimensional Landau levels. Solid State Commun. {\bf 109}, 389 (1999).

\bibitem{E} M. Shayegan, H. C. Manoharan, S. J. Papadakis, and E. P. DePoortere, Anisotropic transport of two-dimensional holes in high Landau levels, Physica E {\bf 6}, 40 (2000).

\bibitem{F} E. Fradkin, S. A. Kivelson, M. J. Lawler, J. P. Eisenstein, and A. P. Mackenzie, Nematic Fermi fluids in condensed matter physics. Annu. Rev. Condens. Matter Phys. {\bf 1}, 153 (2010).

\bibitem{NC1} J. P. Eisenstein, L. N. Pfeiffer, and K. W. West, Compressibility of the two-dimensional electron gas: Measurements of the zero-field exchange energy and fractional quantum Hall gap. Phys. Rev. B {\bf 50}, 1760 (1994).

\bibitem{NC2} X. Ying, S. R. Parihar, H. C. Manoharan, and M. Shayegan, Quantitative determination of many-body-induced interlayer charge transfer in bilayer electron systems via Shubnikov-de Haas measurements. Phys. Rev. B {\bf 52}, R11611(R) (1995).

\bibitem{ferro} E. P. De Poortere, E. Tutuc, S. J. Papadakis, and M. Shayegan, Resistance spikes at transitions between quantum Hall ferromagnets. Science {\bf 290}, 1546 (2000).

\bibitem{graph1} Y. Cao, V. Fatemi, S. Fang, K. Watanabe, T. Taniguchi, E. Kaxiras, and P. Jarillo-Herrero, Unconventional superconductivity in magic-angle graphene superlattices. Nature {\bf 556}, 43 (2018).

\bibitem{graph2} Y. Cao, V. Fatemi, A. Demir, S. Fang, S. L. Tomarken, J. Y. Luo, J. D. Sanchez-Yamagishi, K. Watanabe, T. Taniguchi, E. Kaxiras, R. C. Ashoori, and P. Jarillo-Herrero, Correlated insulator behaviour at half-filling in magic-angle graphene superlattices. Nature {\bf 556}, 80 (2018).

\bibitem{modulationdoping} R. Dingle, H. L. St\"{o}rmer, A. C. Gossard, and W. Wiegmann, Electron mobilities in modulation-doped semiconductor heterojunction superlattices. Appl. Phys. Lett. {\bf 33}, 665 (1978).

\bibitem{Ga1} G. C. Gardner, S. Fallahi, J. D. Watson, and M. J. Manfra, Modified MBE hardware and techniques and role of gallium purity for attainment of two dimensional electron gas mobility $>35\times10^6$ cm$^2$/V s in GaAs/AlGaAs quantum wells grown by MBE. J. Cryst. Growth {\bf 441}, 71 (2016).

\bibitem{Ga2} F. Schlapfer, W. Dietsche, C. Reichl, S. Faelt, and W. Wegscheider, Photoluminescence and the gallium problem for highest-mobility GaAs/AlGaAs-based 2d electron gases. J. Cryst. Growth {\bf 442}, 114 (2016).

\bibitem{Al1} Y. J. Chung, K. W. Baldwin, K. W. West, M. Shayegan, and L. N. Pfeiffer, Surface segregation and the Al problem in GaAs quantum wells. Phys. Rev. Materials {\bf 2}, 034006 (2018).

\bibitem{Al2} Y. J. Chung, K. A. Villegas Rosales, H. Deng, K. W. Baldwin, K. W. West, M. Shayegan, and L. N. Pfeiffer, Multivalley two-dimensional electron system in an AlAs quantum well with mobility exceeding $2\times10^6$ cm$^2$ V$^{-1}$ s$^{-1}$. Phys. Rev. Materials {\bf 2}, 071001(R) (2018).


\bibitem{labnotes} L. N. Pfeiffer, K. W. West, and K. W. Baldwin, AT\&T Bell Laboratories lab notes, GaAs {\bf 13}, page 80, sample 10-29-91-1 (1991); unpublished.

\bibitem{Ruidu} R. R. Du, H. L. St\"{o}rmer, D. C. Tsui, L. N. Pfeiffer, and K. W. West, Experimental evidence for new particles in the fractional quantum Hall effect. Phys. Rev. Lett. {\bf 70}, 2994 (1993).

\bibitem{Eisenstein} J. P. Eisenstein, K. B. Cooper, L. N. Pfeiffer, and K. W. West, Insulating and fractional quantum Hall states in the first excited Landau level. Phys. Rev. Lett. {\bf 88}, 076801 (2002).

\bibitem{DW1} M. Sammon, M. A. Zudov, and B. I. Shklovskii, Mobility and quantum mobility of modern GaAs/AlGaAs heterostructures. Phys. Rev. Materials {\bf 2}, 064604 (2018).

\bibitem{DW2} M. Sammon, T. Chen, and B. I. Shklovskii, Excess electron screening of remote donors and mobility in modern GaAs/AlGaAs heterostructures. Phys. Rev. Materials {\bf 2}, 104001 (2018).

\bibitem{flatband} For the sake of simplicity, the schematic conduction bands are drawn flat here in the doped region. In reality the bands should be bent to reflect the presence of electric fields that are generated from the dopants and surface charge states. 

\bibitem{fnote1} We take the capacitive energy term ($E_{cap}$) to be much larger than $E_0$ and $E_F$ in the main quantum well. Formally, $\Delta E=E_{cap}+E_0+E_F$, but we are assuming $\Delta E\simeq E_{cap}$, where $E_{cap}$ is related to the 2DES density by $n\simeq2(E_{cap}{\epsilon_b}/{se^2})$. In our work, the standard modulation doped GaAs/AlGaAs heterostructures are designed to work after light illumination, so the donor energy level of Si is taken to be the hydrogenic value of $\simeq6$ meV below the conduction-band edge. In the case of structures working in the dark, the donor energy level would be significantly deeper, pinned by the DX effect. See Ref. [18] for more details. 


\bibitem{Davies} J. H. Davies, \textit{The Physics of Low Dimensional Semiconductors}. (Cambridge University Press, 1997).

\bibitem{Design} Y. J. Chung, K. W. Baldwin, K. W. West, D. Kamburov, M. Shayegan, and L. N. Pfeiffer, Design rules for modulation-doped AlAs quantum wells. Phys. Rev. Materials {\bf 1}, 021002(R) (2017).

\bibitem{newerfootnote} The quasi-Fermi level is used for convenience to describe the electron transfer process and denotes what the Fermi level would be in the doped region of a DWS before equilibrating with the rest of the structure.

\bibitem{footnote} The density-of-states effective mass for the AlAs cladding wells is $\simeq0.46$ (in units of the free electron mass), making the density of states $\simeq6.9$ times larger than GaAs ($m^*=0.067$). This implies that for every electron that goes to the main GaAs quantum well, effectively 6.9 additional electrons must go to the cladding well. Since there are two cladding wells on each side of the main quantum well, the electron yield per Si atom is estimated to be $0.008\times6.9\times4\simeq0.22$.

\bibitem{delta1} M. Santos, T. Sajoto, A. Zrenner, and M. Shayegan, Effect of substrate temperature on migration of Si in planar-doped GaAs. Appl. Phys. Lett. {\bf 53}, 2504 (1988). Erratum, Appl. Phys. Lett. {\bf 55}, 603 (1989).

\bibitem{delta2} A. Zrenner, F. Koch, R. L. Williams, R. A. Stradling, K. Ploog and G. Weimann, Saturation of the free-electron concentration in delta-doped GaAs: the DX centre in two dimensions. Semicond. Sci. Technol. {\bf 3}, 1203 (1988).

\bibitem{secondsub} H. L. St\"{o}rmer, A. C. Gossard, and W. Wiegmann, Observation of intersubband scattering in a 2-dimensional electron system. Solid State Commun. {\bf 41}, 707 (1982).

\bibitem{newfootnote} The red line shown in Fig. 3(b) is calculated by assuming that the density is purely determined by the capacitive term in the equation $n=2(E_{cap}{\epsilon_b}/{se^2})$+ $E_{0,main}+E_F$ so that $n\simeq2(E_{cap}{\epsilon_b}/{se^2})$, and that $E_{cap}=E_{0,cladding}+\Delta E_{C,GaAs-AlAs}$. Here, $\epsilon_b$, $s$, $e$, $E_{0,main}$, $E_F$, $E_{0,cladding}$, and $\Delta E_{C,GaAs-AlAs}$ denote the dielectric constant of the barrier, spacer thickness, electron charge, ground state energy of the main quantum well, Fermi energy with respect to the ground state of the main quantum well, ground state energy of the cladding well, and the conduction-band offset between GaAs($\Gamma$) and AlAs(X), respectively. In calculating the ground state energy of the cladding well, it was further assumed that the cladding wells have symmetric finite potential barriers of $\simeq0.32$ eV, which corresponds to a GaAs-AlAs-GaAs structure. For the actual structure, one side of the quantum well has a lower potential difference of $\simeq0.27$ eV because the material is Al$_{0.24}$Ga$_{0.76}$As instead of pure GaAs. 

\bibitem{powerlaw} M. Shayegan, V. J. Goldman, C. Jiang, T. Sajoto, and M. Santos, Growth of low-density two-dimensional electron system with very high mobility by molecular beam epitaxy. Appl. Phys. Lett. {\bf 52}, 1086 (1988).

\bibitem{Adachi} S. Adachi, GaAs, AlAs, and Al$_x$Ga$_{1-x}$As: Material parameters for use in research and device applications. J. Appl. Phys. {\bf 58}, R1 (1985).

\bibitem{Gabor} V. Shingla, E. Kleinbaum, A. Kumar, L. N. Pfeiffer, K. W. West, and G. A. Cs\'{a}thy, Finite-temperature behavior in the second Landau level of the two-dimensional electron gas. Phys. Rev. B {\bf 97}, 241105(R) (2018).

\bibitem{Manfra} M. Samani, A. V. Rossokhaty, E. Sajadi, S. L\"{u}scher, J. A. Folk, J. D. Watson, G. C. Gardner, and M. J. Manfra, Low-temperature illumination and annealing of ultrahigh quality quantum wells. Phys. Rev. B {\bf 90}, 121405(R) (2014).
 


\end{thebibliography}
\end{document}